\documentclass[fleqn,twoside]{article}
\usepackage{espcrc2}

\usepackage{graphicx}
\usepackage[figuresright]{rotating}

\newcommand{\AmS}{{\protect\the\textfont2
  A\kern-.1667em\lower.5ex\hbox{M}\kern-.125emS}}

\title{Simulations from a new neutrino event generator}

\author{Cezary Juszczak \address[IFT]{Institute of Theoretical Physics, Wroc\l aw University.\\
pl. M. Borna 9, 50-204 Wroc\l aw, Poland}\thanks{The authors were
supported by KBN grant 105/E-344/SPB/ICARUS/P-03/DZ211/2003-2005,
and Jaros\l aw Nowak was also supported by grant IFT/W/2525} ,
Jaros\l aw A. Nowak\addressmark[IFT],
Jan~T.~Sobczyk\addressmark[IFT]}

\begin{document}

\begin{abstract}
We construct a new Monte Carlo generator of events for neutrino
interactions. The dynamical models for quasi-elastic reactions,
$\Delta$ excitation and more inelastic events described by the DIS
formalism with the PDFs modified according to recent JLab data are
used. We describe in detail single pion production channels, which
combine the $\Delta$ excitation and DIS contribution. Many
comparisons of the outcome of simulations with experimental data are
presented. \vspace{1pc}
\end{abstract}

\maketitle

\section{INTRODUCTION}

We present a new Monte Carlo generator of events for neutrino
interactions. The original motivation for our work was to improve
NUX+FLUKA scheme where no separate resonance contribution is present
\cite{NF}. The aim of NUX+FLUKA was to describe interactions of
neutrinos of higher energies and from that point of view the
resonance part was of minor importance. Usually MC generators
contain a resonance contribution described by means of Rein-Sehgal
model covering the kinematical region of hadronic invariant mass
$M+m_{\pi}\leq W\leq 2~GeV$. If for neutrino reactions the
quark-hadron duality holds true  one can assume that contributions
from higher resonances are averaged by deep inelastic scattering
(DIS) structure functions and that only the dominant $\Delta$
resonance has to be treated separately.

The current version of the generator includes various dynamical
models: quasi-elastic \cite{LS}, $\Delta$ excitation \cite{Paschos},
and DIS for which we use GRV94 Parton Distribution Functions (PDF)
\cite{grv94} with modifications proposed by Bodek and Yang
\cite{bodek}. The total cross section for the neutrino scattering is
assumed to be the incoherent sum:
\begin{equation}
\sigma_{total} = \sigma^{CC}_{QE}+\sigma^{CC}_{SPP} +
\sigma^{CC}_{DIS} + \sigma^{NC}_{QE}+\sigma^{NC}_{SPP} +
\sigma^{NC}_{DIS},
\end{equation}
where $\sigma_{SPP}$ is the sum of cross sections for single pion
production (SPP) and CC and NC  denote charge and neutral current
reactions respectively.

The MC generator is organized around the event structure which
contains three vectors of particles: incoming, temporary and
outgoing. It also contains a structure with all the parameters used
and a set of boolean flags tagging the event as QEL, DIS, CC, NC
etc. The input parameters are read at startup from a text file and
the events are stored in the ROOT tree file to simplify further
analysis. Different interactions are implemented as functions acting
on the event structure reading the incoming particles and producing
the temporary ones. The type of the interaction is chosen according
to the ratio of the total cross sections.

In our presentation we focus on single pion channels. We present
many comparisons with the existing experimental data. In the near
future the generator will be supplemented with a module with nuclear
effects.

\section{FRAGMENTATION ALGORITHM}

\begin{figure}[t]
\vspace{9pt}\label{Pn}
\includegraphics[scale=.32, angle=270]{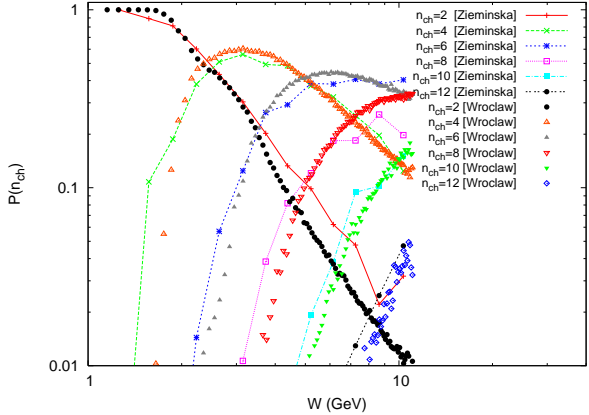}
\vspace*{-0.5cm}
 \caption{Charged hadrons multiplicities for $\nu p
\to \mu^- X^{++}$. Data points taken from \cite{Zieminska} are
connected by lines. The results of our simulations are shown as
separated points. }
\end{figure}

\begin{figure}[htb]\label{fig:spp_vs_el}
\includegraphics[scale=.6]{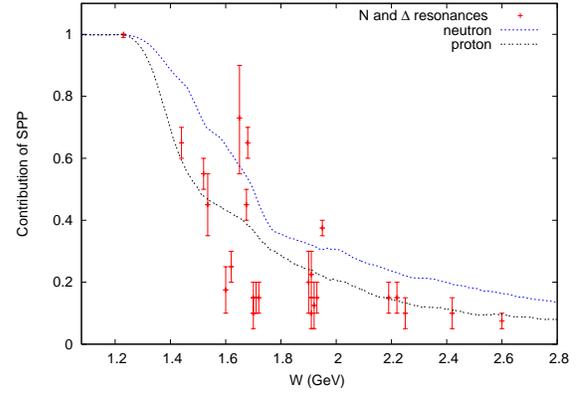}
\vspace*{-0.5cm}
 \caption{Comparison of CC 1-pion functions with
elasticities of $N$ and $\Delta$ resonances. For neutron the sum of
1-pion functions for two exclusive channels is shown.}
\end{figure}

\begin{figure}\label{nu_N_mu_X}
\includegraphics[scale=.6]{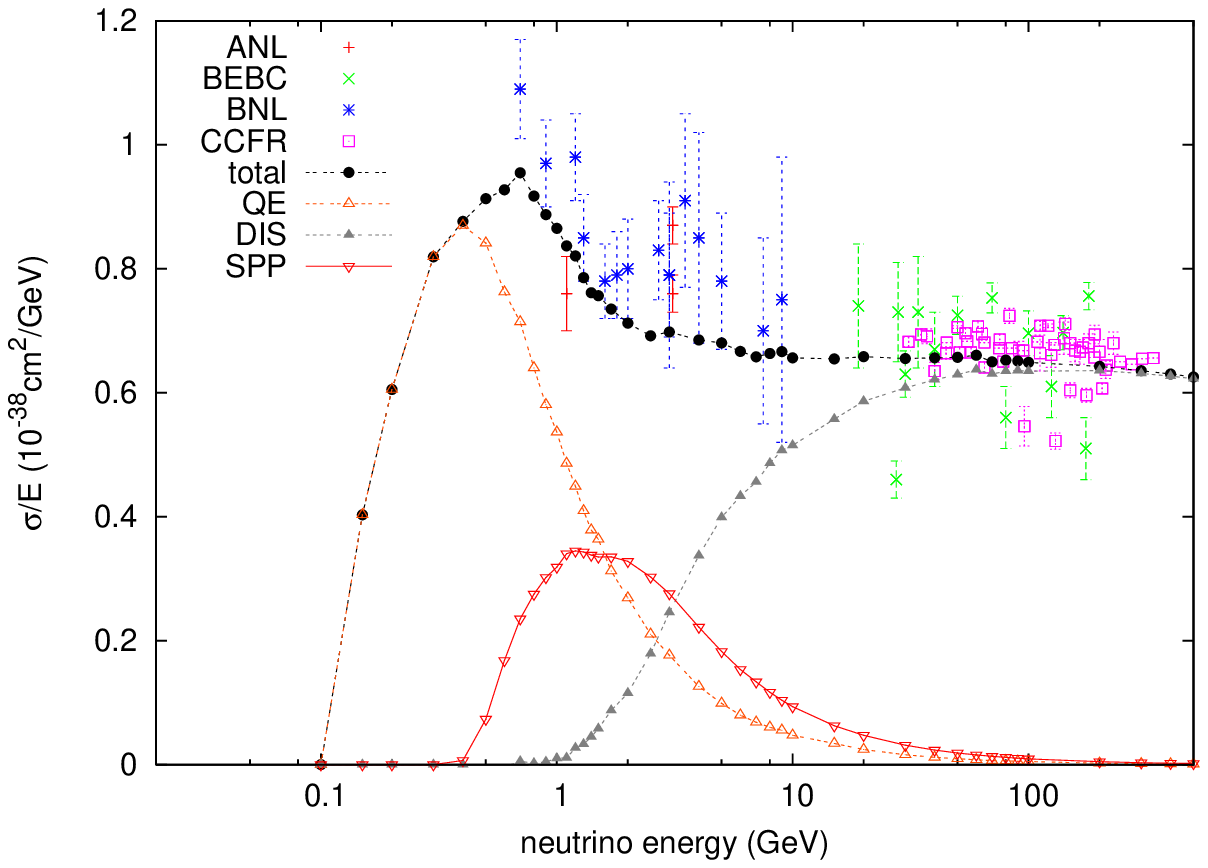}
\vspace*{-0.5cm}
 \caption{CC cross section for $\nu N \to \mu^- X$
interaction. Total cross section is split into contributions from
quasi-elastic, SPP ($W<2$ GeV) and more inelastic processes. Data
points are taken from \cite{ANL}, \cite{BEBC}, \cite{BNL},
\cite{kitagaki}, \cite{CCFR}}
\end{figure}

\begin{figure}\label{nubar_N_mu_X}
\includegraphics[scale=.6]{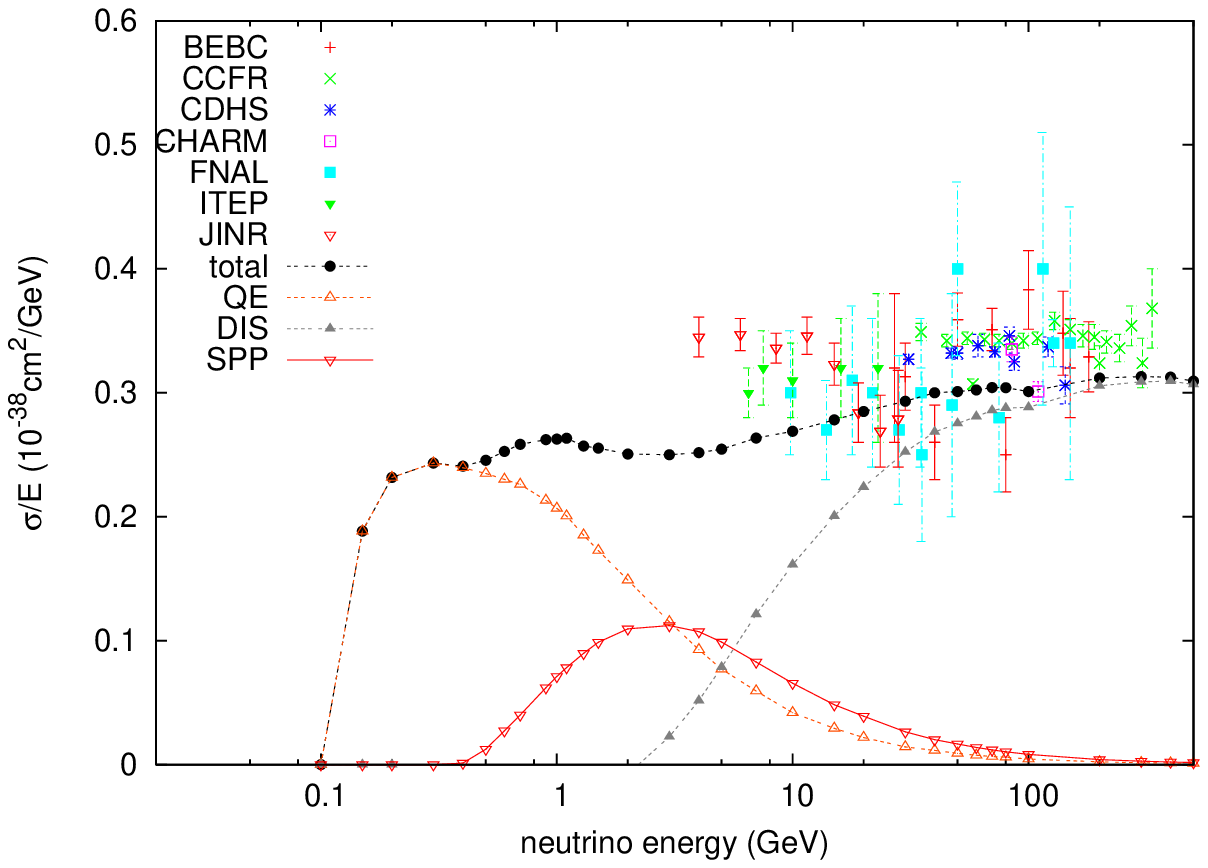}
\vspace*{-0.5cm} \caption{CC cross section for $\bar \nu N \to \mu^+
X$ interaction. Total cross section is split into contributions from
quasi-elastic, SPP ($W<2$ GeV) and more inelastic processes. Data
points are taken from \cite{CCFR}, \cite{CDHS}, \cite{BEBC},
\cite{CHARM}, \cite{FNAL}, \cite{ITEP}, \cite{JINR}}
\end{figure}

In our MC we use the DIS formalism to generate events in the whole
kinematical region where inelastic reactions are possible. In order
to get the event record for the final state we assume that
interaction occurs always on a particular parton  and then the
fragmentation of interacting quark and spectator is performed by
means of PYTHIA6 \cite{pythia} routines.

The inclusive cross section for the  scattering off nucleon is given
by
\begin{eqnarray}
\frac{{d^2 \sigma ^{\nu (\bar{\nu})} (E)}}{{dxdy}} = \frac{{G_F^2 ME}}{\pi }\left[ {\left( {xy^2  + \frac{{ym^2 }}{{2ME}}} \right)F_1 \left( {x,Q^2 } \right)} \right. \hfill \nonumber\\
+ \left( {1 - y - \frac{{Mxy}}{{2E}} - \left( {\frac{m}{{2E}}}
\right)^2  - \frac{{m^2 }}{{2MEx}}} \right)F_2 \left( {x,Q^2 }
\right) \hfill \nonumber
\end{eqnarray}
\begin{equation}\label{DIS}
\ \ \ \ \ \ \ \  \left. { \pm \left( {xy - \frac{{xy^2 }}{2} -
\frac{{ym^2 }}{{4ME}}} \right)F_3} \left( {x,Q^2 }\right)  \right].
\end{equation}

Structure functions are assumed to be those defined in the parton
model i.e. the combinations of a PDFs
\begin{eqnarray}\label{FS}
F_1 \left( {x,Q^2 } \right) &=& \sum\limits_j {\left[ {q_j \left( {x,Q^2 } \right) + \bar q_j \left( {x,Q^2 } \right)} \right]} \nonumber \\
F_3 \left( {x,Q^2 } \right) &=& 2\sum\limits_j {\left[ {q_j \left( {x,Q^2 } \right) - \bar q_j \left( {x,Q^2 } \right)} \right]}  \\
F_2 \left( {x,Q^2 } \right) &=& 2xF_1 \left( {x,Q^2 } \right)
\nonumber
\end{eqnarray}
Using structure functions (\ref{FS}) the cross section (\ref{DIS})
is rewritten in terms of contributions from separate partons $q_i$

\begin{equation}
\frac{{d^2 \sigma ^{\nu q_i\to \mu q_j} }} {{dx dy}} \sim q_i K_i,
\end{equation}
where $K_i$ is a kinematic factor for parton $q_i$.\\

A probability of reaction on a given quark is:
\begin{equation}
P(q_i) = \frac{d^2\sigma^{q_i}/dxdy }{\sum
\limits_i{{d^2\sigma^{q_i}}}/dxdy}
\end{equation}

PYTHIA fragmentation routines require a system of quark and diquark
and perform fragmentation and hadronization using the LUND
algorithm. Depending on the interacting parton, we distinguish
several cases \cite{Sartogo}:
\begin{itemize}
    \item In the case of the scattering off the  valence quark, a string is formed from the
    created quark and the remaining diquark.
    \item In the case of scattering off a sea quark u or d, the remaining anti-quark
    annihilates with appropriate valence quark,
    and the created quark forms a string with the the remaining diquark, exactly as in the previous case.
    \item If scattering off an anti-quark u gives an anti-quark d or scattering off an anti-quark d gives
    an anti-quark u, the created parton annihilates with a valence quark.
    \item If scattering off an anti-quark u gives a strange anti-quark s or scattering off
    an anti-quark d gives an anti-quark c, it creates
    with one of valence quarks a strange or a charm meson and the remaining quarks form a string for the fragmentation.
    \item In the cases of scattering off a strange quark or anti-quark, the remaining strange
    constituent creates a strange meson with one of valence quarks and the remaining quarks form a string
    for the fragmentation.
\end{itemize}

We fine tuned the PYTHIA6 generator parameters of the fragmentation.
In fig. (1) the comparison of the charged particles multiplicities

\begin{equation}\label{Pn_eq}
    P(n_{ch}) = \sigma(n_{ch})/\sum \limits_{n_{ch}} \sigma(n_{ch})
\end{equation}
as obtained  from our simulation with the data from the Fermilab
bubble chamber \cite{Zieminska} is shown.

\begin{figure}\label{fig:total_cc_nu_proton_pionplus}
\includegraphics[scale=.6]{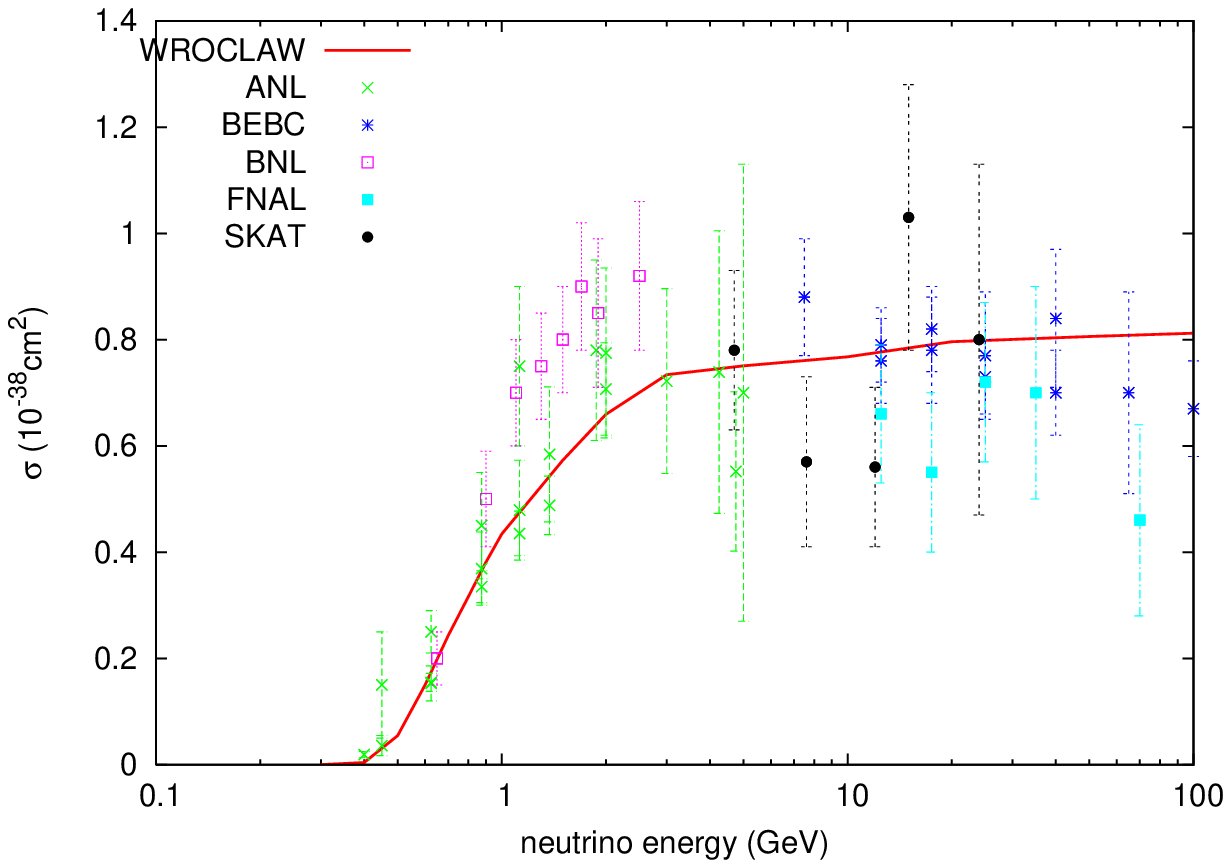}
\caption{Cross section for $\nu p \to \mu^- \pi^+ p$. For data
points and for simulations only events with hadronic
 mass $W<2$ GeV were included. Data points are taken from \cite{ANL}, \cite{BEBC},
 \cite{BNL}, \cite{kitagaki}, \cite{FNAL}, \cite{SKAT}}
\end{figure}

\begin{figure}\label{fig:total_cc_nu_neutron_pionplus}
\includegraphics[scale=.6]{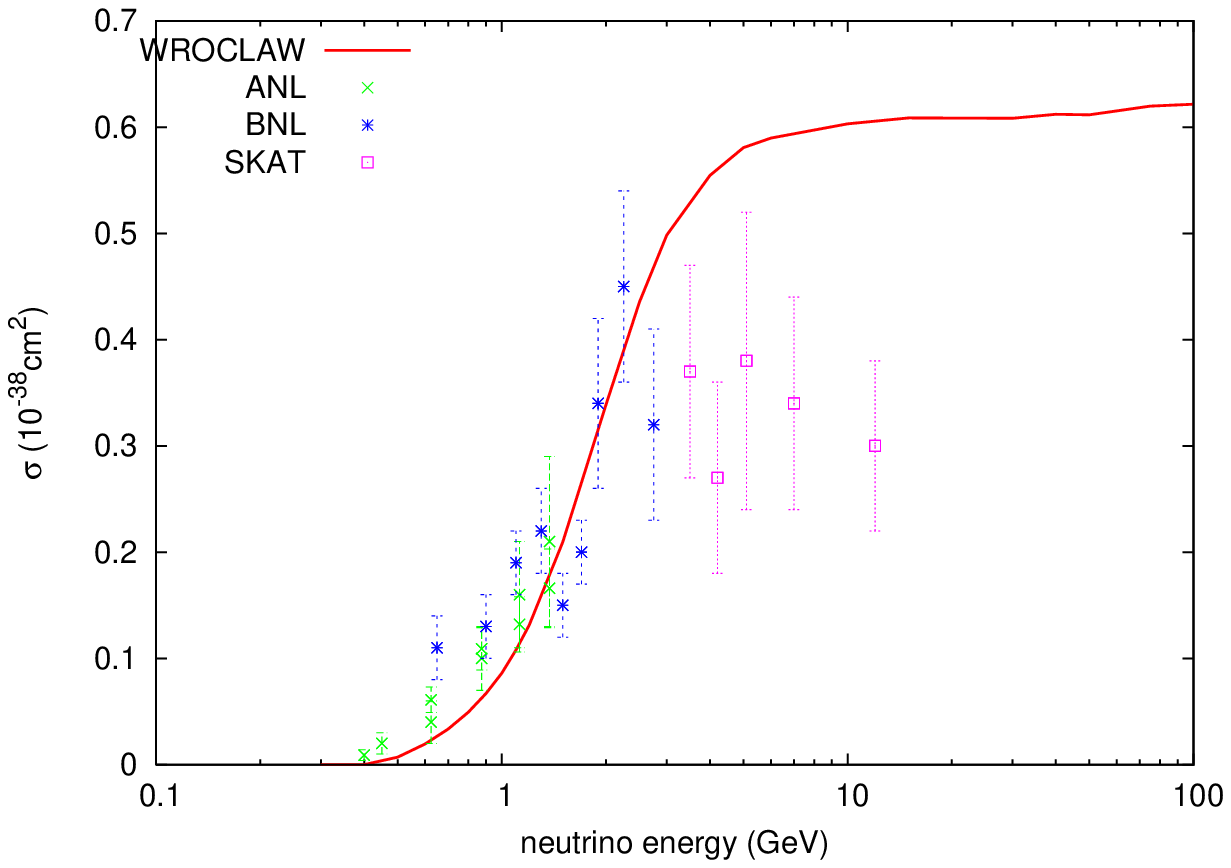}
\caption{Cross section for $\nu n \to \mu^- \pi^+ n$. For data
points and for simulations only events with hadronic
 mass $W<2$ GeV were included. Data points are taken from \cite{ANL}, \cite{BNL}, \cite{kitagaki}, \cite{SKAT}}
\end{figure}

\begin{figure}\label{fig:total_cc_nu_neutron_pionzero}
\includegraphics[scale=.6]{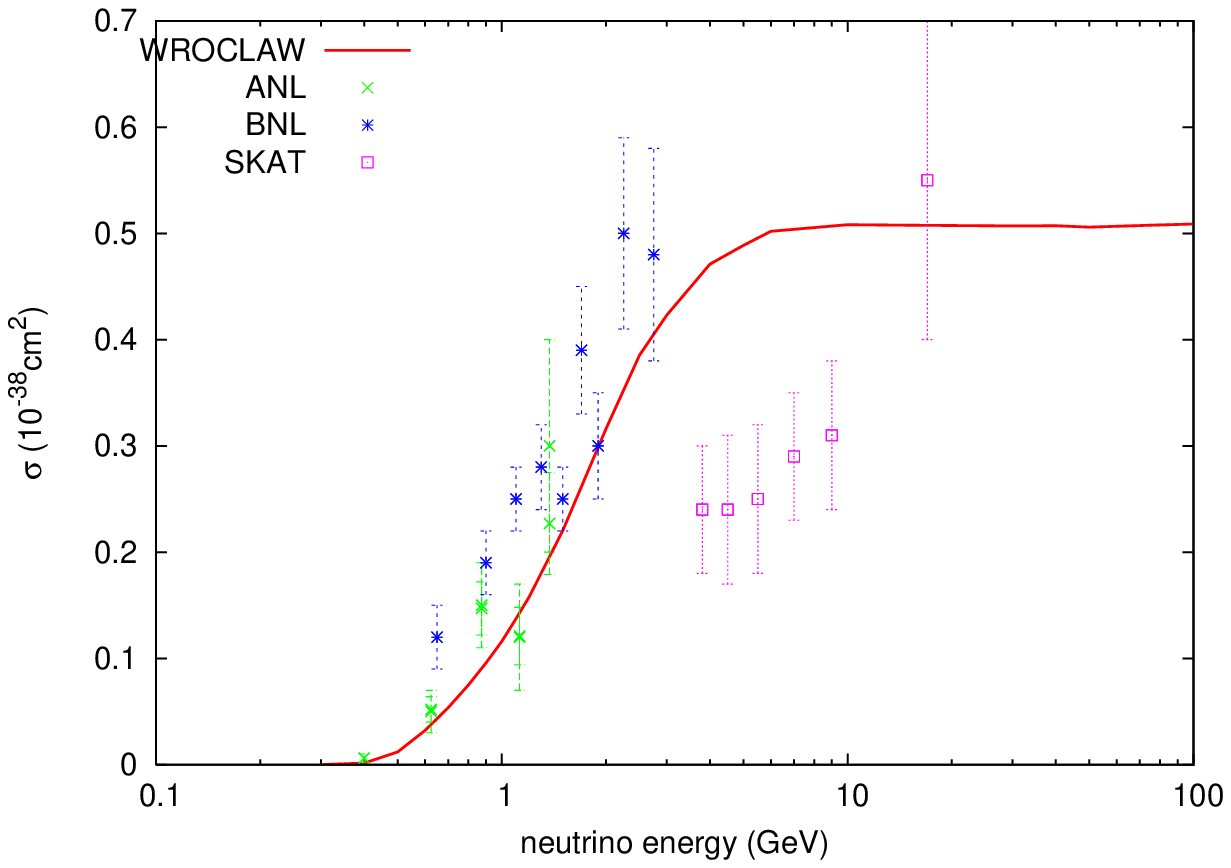}
\caption{Cross section for $\nu n \to \mu^- \pi^0 p$. For data
points and for simulations only events with hadronic mass $W<2$ GeV
were included. Data points are taken from \cite{ANL}, \cite{BNL},
\cite{kitagaki}, \cite{SKAT}}
\end{figure}

\section {1-PION FUNCTIONS}


The only resonance we consider is the $\Delta$ and we have to
estimate the single pion production cross section as a fraction of
the inclusive DIS cross section extrapolated into the resonance
region. This is done separately for each SPP channel and the
obtained fractions are called 1-pion functions. They are the
probabilities that in a given point in the kinematically allowed
region the final state is that of SPP.

\begin{equation}
f^{SPP}(W,\nu )=\frac{\displaystyle
   \frac{
   d^2\sigma^{DIS-SPP}
        }{
        dWd\nu
        }          }{\displaystyle
    \frac{
    d^2\sigma^{DIS}
         }{
         dWd\nu
         }
                      }
\end{equation}

In our generator $f^{SPP}$ are reconstructed using the LUND
fragmentation algorithm. They turn out to be functions of $W$ only
and are shown in fig. (2). We see that up to the threshold for two
pion production, 1-pion function for proton and the sum of functions
for neutron are equal 1. In more common langauge 1-pion functions
can be recognized as average elasticities of resonances
$\Gamma_{N\pi}/ \Gamma_{total}$ \cite{Olga1}. In fig. (2) we see
that in fact the values of 1-pion functions are close to resonance
elasticities in a wide range of hadronic invariant mass.

\begin{figure}\label{fig:total_nc_nu_neutron_pionplus}
\includegraphics[scale=.6]{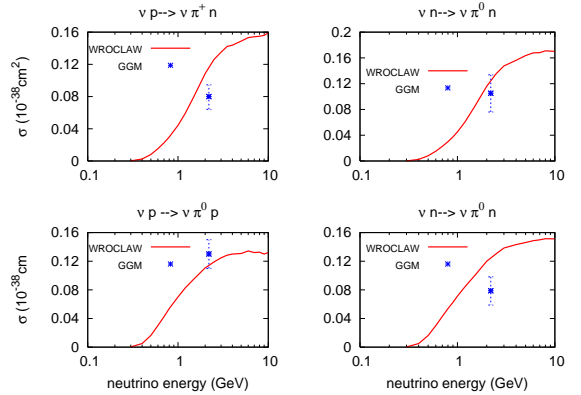}
\caption{Cross section for NC SPP channels. For data points and for
simulations only events with hadronic mass $W<2$ GeV were included.
Data points are taken from \cite{GGM_NC}}
\end{figure}

\begin{figure}\label{fig:diff_proton_pionplus}
\includegraphics[scale=.3, angle = 270]{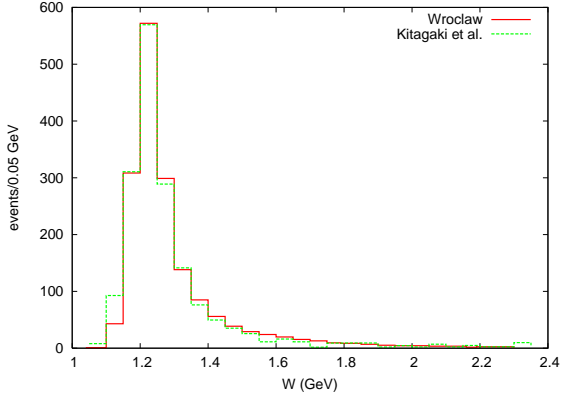}
\vspace*{-0.5cm}
  \caption{ Distribution of events in hadronic mass for BNL
experiment \cite{kitagaki} and predictions of our Monte Carlo for
$\nu p \to \mu^- \pi^+ p$ }\vspace*{-0.5cm}
\end{figure}

\section {SINGLE PION PRODUCTION}

Our model of SPP combines in a smooth way the $\Delta$ excitation
model with the SPP part of the DIS cross section. We choose a linear
transition with respect to hadronic invariant mass $W \in (1.3,
1.6)$ GeV. As a bonus we obtain an artificial resonance-like
behavior of the cross section at $W\sim 1.5$ GeV which closely
resembles the contribution from the $D_{13}, S_{11}$ resonances
\cite{Olga2}. We describe the non-resonant background as a small
admixture of the DIS SPP contributions at low values of $W$. Our MC
reproduces the following analytical expression for the cross
section:
\begin{eqnarray}
\frac{d\sigma^{SPP}}{dW}&=&\frac{d\sigma^{\Delta}}{dW}\left(1-\alpha(W)\right) \nonumber \\
&+& \frac{d\sigma^{DIS}}{dW}F^{SPP}(W)\alpha(W)
\end{eqnarray}


\begin{figure}\label{fig:diff_neutron_pionplus}
\includegraphics[scale=.3, angle = 270]{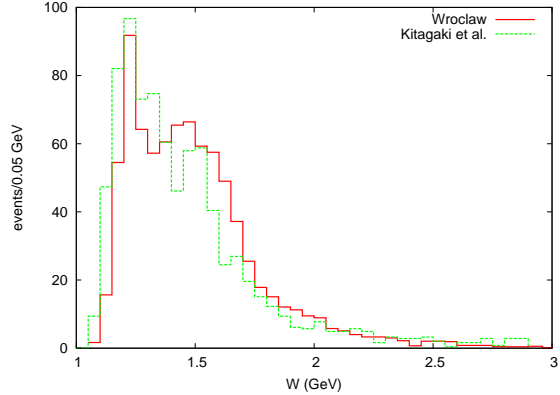}
\vspace*{-0.5cm}
 \caption{ Distribution of events in hadronic mass
for BNL experiment \cite{kitagaki} and predictions of our Monte
Carlo for $\nu n \to \mu^- \pi^+ n$ } \vspace*{-0.5cm}
\end{figure}
where
\begin{eqnarray}
\alpha(W) &=& \Theta(1.3GeV -W)\frac{W-W_{th}}{W_{min} - W_{th}}\alpha_0 \nonumber \\
&+& \Theta(W_{max} -W)\Theta(W-W_{min})\\
& & \frac{W-W_{min}+\alpha_0(W_{max}-W)}{W_{max}-W_{min}} \nonumber
\\
&+& \Theta(W-W_{max}) \nonumber
\end{eqnarray}
$\alpha_0 \in (0, 0.3)$, depending on the channel. We note that a
similar value for the division line in the hadronic invariant mass
between resonance and DIS contributions
($W_{cut}^{DIS}=W_{cut}^{RES}=1.5 \pm 0.02$) was found by Naumov et
al.\cite{Naumov} by fitting procedure to the existing set of
experimental data.

The performance of our generator is presented on a series of plots.
First we show contributions to the inclusive cross section for
neutrino and anti-neutrino interaction on isoscalar target (figs.
3-4) from three theoretically separated dynamical mechanism. The SPP
contribution is restricted by a cut $W\leq 2$ GeV.

\begin{figure}\label{fig:diff_neutron_pionzero}
\includegraphics[scale=.3, angle = 270]{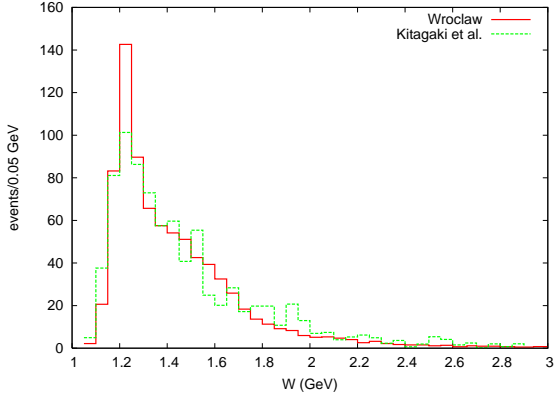}
\vspace*{-0.5cm}
 \caption{ Distribution of events in hadronic mass
for BNL experiment \cite{kitagaki} and predictions of our Monte
Carlo for $\nu n \to \mu^- \pi^0 p$ } \vspace*{-0.5cm}
\end{figure}

The cross sections for CC SPP channels are shown in figs. 5-7.  We
conclude that the agreement with the data is satisfactory. In fig. 8
we show the plots for NC SPP channels.

We also compared the distribution of events in hadronic mass for SPP
channels with the data from the BNL experiment. It is an important
test because our procedure of modelling SPP channels is different
from what is done in other MC codes. We used the BNL neutrino beam
and generated the same number of events as reported in
\cite{kitagaki}. The results are shown in figs. (8-10). In the case
of neutrino-proton reaction the agreement is excellent. In the case
of $\nu n \to \mu^- \pi^0 p$ reaction the agreement is very good but
our simulations give too high $\Delta$ peak. In the case of $\nu n
\to \mu^- \pi^+ n$ there is an experimentally measured access of
events with small invariant mass (smaller then $M_{\Delta}=1232$
MeV) which is not reproduced by our simulations.

\section{FINAL REMARKS}

We find our results for SPP encouraging. An improvement in the $\nu
n \to \mu^- \pi^+ n$ channel can probably be achieved by a more
accurate treatment of the non-resonant background.\\

\section*{Acknowledgments}

The collaboration of Krzysztof Graczyk at early stages of the MC
generator and many useful discussions are acknowledged with
pleasure.

\end{document}